\documentclass[aps,pre,twocolumn,groupedaddress,showpacs,floatfix]{revtex4}
\usepackage{dcolumn}
\usepackage{amsmath}
\usepackage{amssymb}
\usepackage{graphicx}
\usepackage{bm}
\usepackage[T1]{fontenc}
\usepackage{color}
\usepackage{url}
\usepackage[bf]{subfigure}
\usepackage{rotating}

\begin{document}
\title{Resilience of networks to environmental stress: From regular to random networks}
\author{Young-Ho Eom}
\email{youngho.eom@strath.ac.uk}
\affiliation{Department of Mathematics and Statistics, University of Strathclyde, 26 Richmond Street, Glasgow G1 1XH, United Kingdom}
\affiliation{Departamento de Matem\'aticas, Universidad Carlos III de Madrid, 28911 Legan\'es, Spain}

\begin{abstract}
Despite the huge interest in network resilience to stress, most of the studies have concentrated on internal stress damaging network structure (e.g., node removals). Here we study how networks respond to environmental stress deteriorating their external conditions. We show that, when regular networks  gradually disintegrate as environmental stress increases, disordered networks can suddenly collapse at critical stress with hysteresis and vulnerability to perturbations. We demonstrate that this difference results from a trade-off between node resilience and network resilience to environmental stress. The nodes in the disordered networks can suppress their collapses due to the small-world topology of the networks but eventually collapse all together in return. Our findings indicate that some real networks can be highly resilient against environmental stress to a threshold yet extremely vulnerable to the stress above the threshold because of their small-world topology. 
\end{abstract}

\maketitle
\section{Introduction}
Networks usually maintain their essential functions owing to their resilience although they are under constant stress such as damages, failures, and environmental changes~\cite{Barabasi2016book,Gao2016,Albert2000}. But networks can lose their resilience and collapse when stress exceeds a critical level (i.e., threshold). How networks lose their resilience relies on their topology. Homogeneous networks are fragmented when the fraction of randomly removed nodes or links reaches breakdown thresholds whereas scale-free networks are robust to such removals~\cite{Albert2000,Cohen2000,Callaway2000}. The disintegrating pattern of networks at thresholds also depends on their topology. When nodes fail, abrupt transitions to disintegrated state can arise in interdependent networks but smooth transitions occur in isolated networks~\cite{Buldyrev2010,Parshani2010,Gao2011,Gao2011a,Bashan2013}. Identifying the relation between the response of networks to stress and their topology is a vital task for their stable function.

Stress can cause normally two types of effects on networks. Accordingly one can define two types of stress. Internal stress damages network structure and causes changes in the weighted adjacency matrix of networks (e.g., loss of nodes, links, or coupling strength), whereas environmental stress deteriorates external conditions of networks (e.g., changes in external parameters determining the state of nodes). Both types of stress can have profound impact on networks. In a network of patchy ecosystems, for instance, not only loss of patches but also climate changes and diminishing resources can be serious threats. In the Internet or power grids, growing demand in the systems can cause the systemic collapse as the local failures can. However, the role of network topology in network resilience to environmental stress has been neglected, in contrast to the great attention to internal stress~\cite{Gao2016,Albert2000,Cohen2000,Callaway2000,Cohen2001,Vazquez2003,Valente2004,Gallos2005,Buldyrev2010,Parshani2010,Gao2011,Gao2011a,Bashan2013}, despite existing works on networks under environmental changes~\cite{Lozano2007,Saavedra2013,Dakos2014,Lever2014}. For example, recent studies on mutualistic networks of interacting species showed how such networks behave when interaction strength~\cite{Saavedra2013,Dakos2014} or growth rate~\cite{Lever2014} are reduced by environmental changes. The role of their nested network topology was also highlighted~\cite{Lever2014}. But these works have focused on mutualistic networks, a specific type of ecological networks. How other, possibly more general, networked systems respond to environmental stress remains an open question.

To address this, we consider a simple network model where all nodes are subject to a common external condition. Each node, when isolated, can change abruptly its state (e.g., from active to collapsed) via saddle-node bifurcation if the condition crosses its own threshold. But the coupling among the nodes can affect their collective response to the condition. This model is motivated by the models for ecosystems exposed to environmental stress. Ecosystems can undergo sudden state changes when environmental stress exceeds catastrophic thresholds, often called \emph{tipping points}~\cite{Barnosky2012,Scheffer2012,Scheffer2009a}. Recent studies show that tipping points arise in various systems ranging from ecosystems and populations~\cite{Scheffer2001,Carpenter2005,Kefi2007,Scheffer2009,Drake2010,Carpenter2011,Dai2012,Veraat2012} to climate systems~\cite{Lenton2008} and financial markets~\cite{May2008}. To anticipate whether these large-scale systems respond to environmental stress abruptly or gradually, many studies have considered theoretical models composed of connected components which shift their states via bifurcation in response to changing external conditions~\cite{Rietkerk2004,Guttal2009,Fernandez2009,Dakos2010,Dai2013,Weissmann2016,vanNes2005,Martin2015}. In particular, some of these studies reported that abrupt responses to environmental changes can be smoothened if there exist strong heterogeneity in the response of the components, limited coupling among the components, or sufficient noise~\cite{vanNes2005,Martin2015} (But note that a recent study~\cite{Weissmann2017} claimed that there is a bias in the numeric simulations of the Ref.~\cite{Martin2015}.) These models have assumed that their connection topology among the components is perfectly regular, whereas most real networks have disorder in their topology. 

In this work we study the effects of topological disorder on the response of networks to environmental stress. Specifically, we investigate whether topological disorder can alter the nature of response as the heterogeneity or the coupling can. We show that as the external condition deteriorates, regular networks gradually crumble, whereas disordered networks suddenly collapse into distinct states with strong hysteresis and vulnerability to perturbations. This means that tipping points arise in the disordered networks. We demonstrate that the nodes in the disordered networks can keep their functions owing to the networkwide support caused by the small-world topology of the networks but ultimately they collapse all together in return. This indicates that there is a trade-off, determined by network topology, between node resilience and network resilience to environmental stress and real networks may suddenly collapse under environmental stress owing to their small-world topology.

\section{The network model}
\subsection{Nodes: Minimal systems}
\begin{figure}[!b]
\centerline{\includegraphics[width=0.98\linewidth,angle=0]{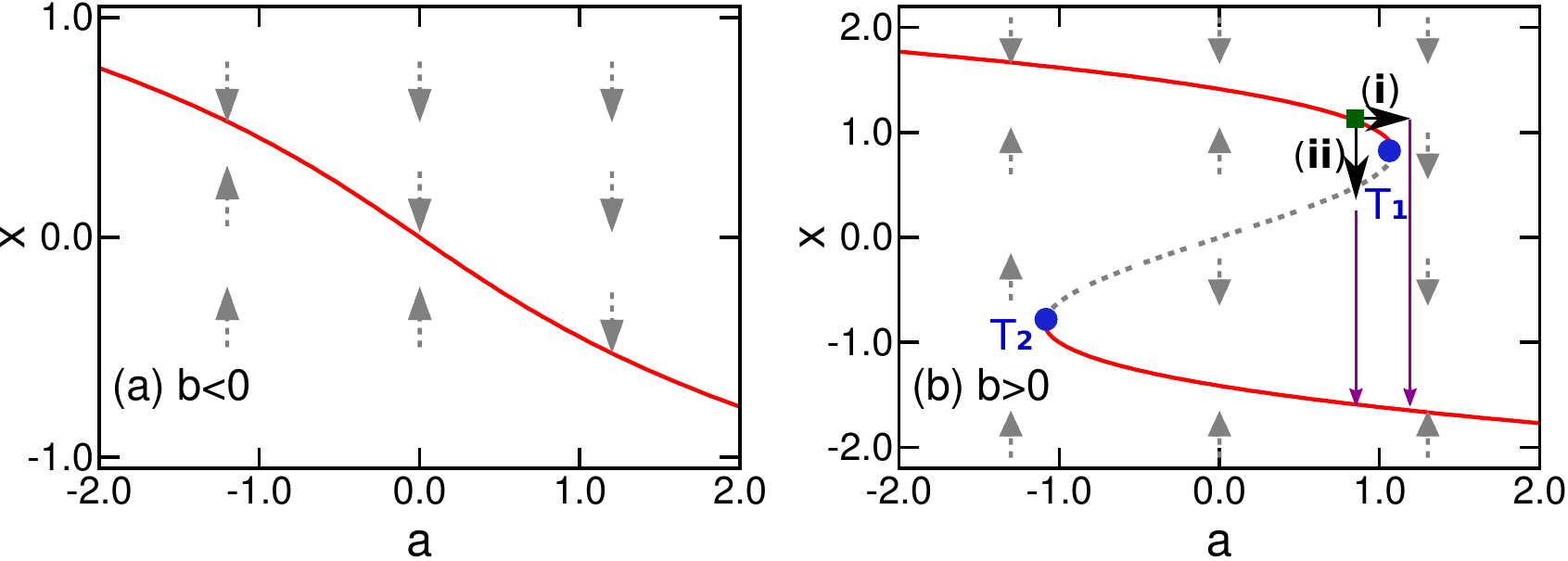}}
\caption{(color online). Bifurcation diagrams of the minimal system whose dynamics is described by Eq.~(\ref{eq:1}) with (a) $b=-2.0$ and (b) $b=2.0$. The red lines represent the stable states and the dotted gray line represents the unstable states. The blue filled circles (i.e., $T_1$ and $T_2$) in (b) denote the tipping points where saddle-node bifurcation occurs. Two ways of catastrophic shifts between distinct stable states are illustrated in (b): (i) the parameter $a$ is passed the thresholds $T_1$ (the horizontal arrow) and (ii) the system's state is crossed the unstable states by perturbations (the black vertical arrow).} \label{fig:1}
\end{figure}

We take a cusp catastrophe system for the dynamics of the nodes in our model. It is a minimal mathematical model for systems changing abruptly their states via saddle-node bifurcation~\cite{Zeeman1976,Strogatz1994}. The state of this minimal system is evolved by the following differential equation
\begin{equation} \label{eq:1}
 \frac{dx}{dt} = -a+bx-{x}^{3},
\end{equation}
where $x$ is the state of the system (e.g., activity). We consider the system as \textit{active} (\textit{collapsed}) for $x>0$ ($x<0$). The parameter $a$ and $b$ denote the environmental stress on the system and the growth rate of the system, respectively. The cubic term $-x^{3}$ is to prevent the state from diverging. This minimal system can be considered as a mean-field model for systems with tipping points because saddle-node bifurcation can correspond to the simplest form of catastrophic shifts~\cite{vanNes2016,Kuehn2011,Brummitt2015}. Figure~\ref{fig:1} represents the bifurcation diagrams of the system. When $b<0$, the equilibrium state curve is gradual (Fig.~\ref{fig:1}(a)). When $b>0$, the equilibrium state curve is folded at the tipping points (i.e., $T_1$ and $T_2$) where one of the stable states disappears via saddle-node bifurcation (Fig.~\ref{fig:1}(b)). If the system is close to a tipping point, then small changes in the stress (the horizontal arrow in Fig.~\ref{fig:1}(b)) or in the state (the black vertical arrow in Fig.~\ref{fig:1}(b)) can radically move the system to another stable state (e.g., from active to collapsed). Once the system collapsed at $T_1$, we should restore the stress to $T_2<T_1$ for the recovery (i.e., hysteresis).

\subsection{Networks of the minimal systems}
The state of node $i$ in a network of $N$ minimal systems is evolved by 
\begin{equation} \label{eq:2}
 \frac{dx_i}{dt} = -a+b_ix_i-{x_i}^{3} + \frac{D}{k_i}\sum_{j=1}^N A_{ij}(x_j-x_i),
\end{equation}
where $x_i$ is the state of node $i$ ($i=1,\ldots,N$). The parameter $a$ is the environmental stress and $b_i$ is the growth rate of node $i$ randomly drawn from a uniform distribution $B$. $D$ is the coupling strength, $k_i$ is the number of connections (i.e., the degree) of node $i$, and $A_{ij}$ is the element of the adjacency matrix of the network which is unity when nodes $i$ and $j$ are connected and zero otherwise. The coupling term denotes the dispersion of activity among the connected nodes~\cite{vanNes2005,Dakos2010}. The network's state $X$ is defined as $X=\frac{1}{N}\sum_i x_i$. The network is active (collapsed) if $X>0$ ($X<0$) as the nodes. We consider networks with the same average degree $\langle k \rangle=(1/N)\sum_i k_i$ and the same size $N=10^4$ generated by three network models: the Watts-Strogatz (WS) model~\cite{Watts1998}, the Erd\H{o}s-R\'{e}yni (ER) model~\cite{Bollobas2001}, and the Barab\'asi-Albert (BA) model~\cite{Barabasi1999}. The WS model allows us to construct networks ranging from regular lattice ($P=0$) to disordered lattice ($P=1$) by tuning a link rewiring probability $P$ (i.e., the topological disorder in the networks). We consider two-dimensional lattices with periodic boundaries for the WS model. The ER model and the BA model generate random networks with Poisson degree distributions and with power-law degree distributions, respectively. 

\section{Results}
Here we aim at checking whether the gradual response of regular networks resulting from their strong heterogeneity or limited coupling can change to an abrupt response when the disorder in network topology increases. Accordingly we choose the parameter values $D=1.0$, $B=[0,3]$, and $\langle k \rangle=4$ in our numeric simulations. Such parameter values generate networks with sufficient heterogeneity (i.e., broad $B$) and limited coupling (i.e., low $D$ and $\langle k \rangle$), so that one can observe gradual response to environmental stress in regular networks (i.e., the WS network with $P \ll 1$). But note that, although networks are perfectly regular, they can respond abruptly to environmental stress if we consider the higher value of D, the narrower $B$, or the higher $\langle k \rangle$ as predicted in the previous studies~\cite{vanNes2005,Martin2015} (see Appendix A for the observed abrupt responses of the regular networks in such cases). Here we did not consider noise to focus on the role of network topology in the response to environmental stress. Thus the applicability of our results is limited to weakly noisy situations because noise can affect the nature of the response~\cite{Martin2015}. We focus on synthetic networks because we can tune the topological disorder. All results are obtained by a fourth-order Runge-Kutta method with a time step of $0.02$.

\begin{figure}[!tpb]
\centerline{\includegraphics[width=0.99\linewidth,angle=0]{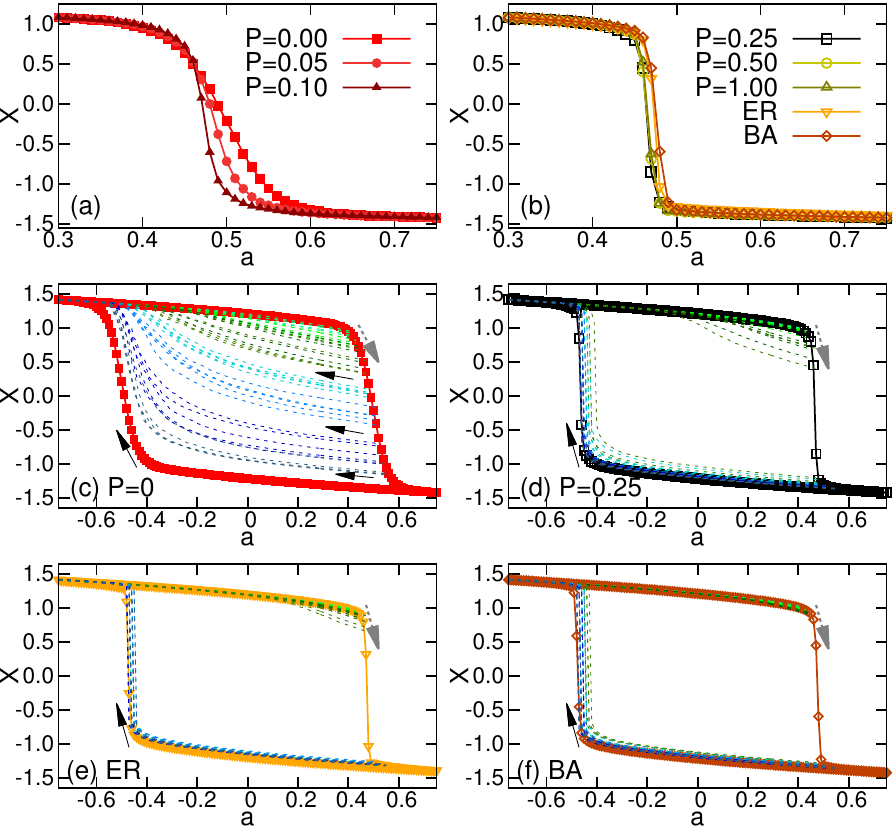}}
\caption{(color online). State diagrams $X(a)$ of the networks of the minimal systems. (a) The Watts-Strogatz (WS) networks with $P\leq0.1$ respond gradually to the increasing stress $a$. (b) The WS networks with $P\geq 0.25$, the Erd\H{o}s-R\'{e}yni (ER), and the Barab\'asi-Albert (BA) respond abruptly to the increasing stress. Increasing and then decreasing the stress, we observed hysteresis in (c) the WS network with $P=0$, (d) the WS network with $P=0.25$, (e) the ER network, and (f) the BA network. Each line with symbols is averaged over 50 realizations. The dotted lines in (c)-(f) show that the hysteresis can be mitigated in the WS network with $P=0$ but not in the WS network with $P=0.25$, the ER network, and the BA network. Each dotted line shows a single realization obtained by increasing the stress $a$ from $a=-0.8$ and then decreasing the stress $a$ when it reaches $a=0.41$, $0.43$, $0.45$, $0.47$, $0.49$, $0.51$, $0.53$, or $0.55$.} \label{fig:2}
\end{figure} 

To identify how these networks respond to the environmental stress, we report the state diagram $X(a)$ of the networks. We consider a \emph{deteriorating} scenario where the stationary value of $X(a)$ is obtained by increasing the stress $a$ with an increment $\delta a=0.01$ from $a=-0.8$ to $a=0.8$. At $a=-0.8$, initially all nodes have the same active state $X_0=2.0$ to obtain the steady state. Subsequently the steady state of the previous step is assigned as the initial condition of the next step. As depicted in Figs.~\ref{fig:2}(a) and \ref{fig:2}(b), all the networks under the deteriorating scenario disintegrate with certain threshold effects. The shape of their state curves, however, varies depending on their topology. The regular networks (i.e., the WS networks with $P \ll 1$) respond gradually to the increasing stress $a$ as expected in the previous works for regular lattices~\cite{vanNes2005,Martin2015}. The state curves, however, become abrupt as $P$ increases (Fig.~\ref{fig:2}(a)). The state curves of the disordered networks (i.e., the WS network with $P\geq 0.25$, the ER network, and the BA network) are order of a magnitude more abrupt than the curve of the WS network with $P=0$ (Fig.~\ref{fig:2}(b)). We also consider an \emph{ameliorating} scenario where the stationary value of $X(a)$ is obtained by decreasing the stress $a$ with a decrement $\delta a=0.01$ from $a=0.8$ to $a=-0.8$ after the deteriorating scenario. Hysteresis is observed in all the networks (Figs.~\ref{fig:2}(c)-\ref{fig:2}(f)). But we can mitigate the hysteresis in the regular networks if we start to decrease the stress before it reaches $a=0.8$ (The dotted lines in Fig.~\ref{fig:2}(c)). No such mitigation is observed in the disordered networks (The dotted lines in Figs.~\ref{fig:2}(d)-\ref{fig:2}(f)). This observation suggests that although the stress exceeds the thresholds, many nodes are still active in the regular networks whereas most nodes immediately collapse in the disordered networks. 

\begin{figure}[!ht]
\centerline{\includegraphics[width=0.98\linewidth,angle=0]{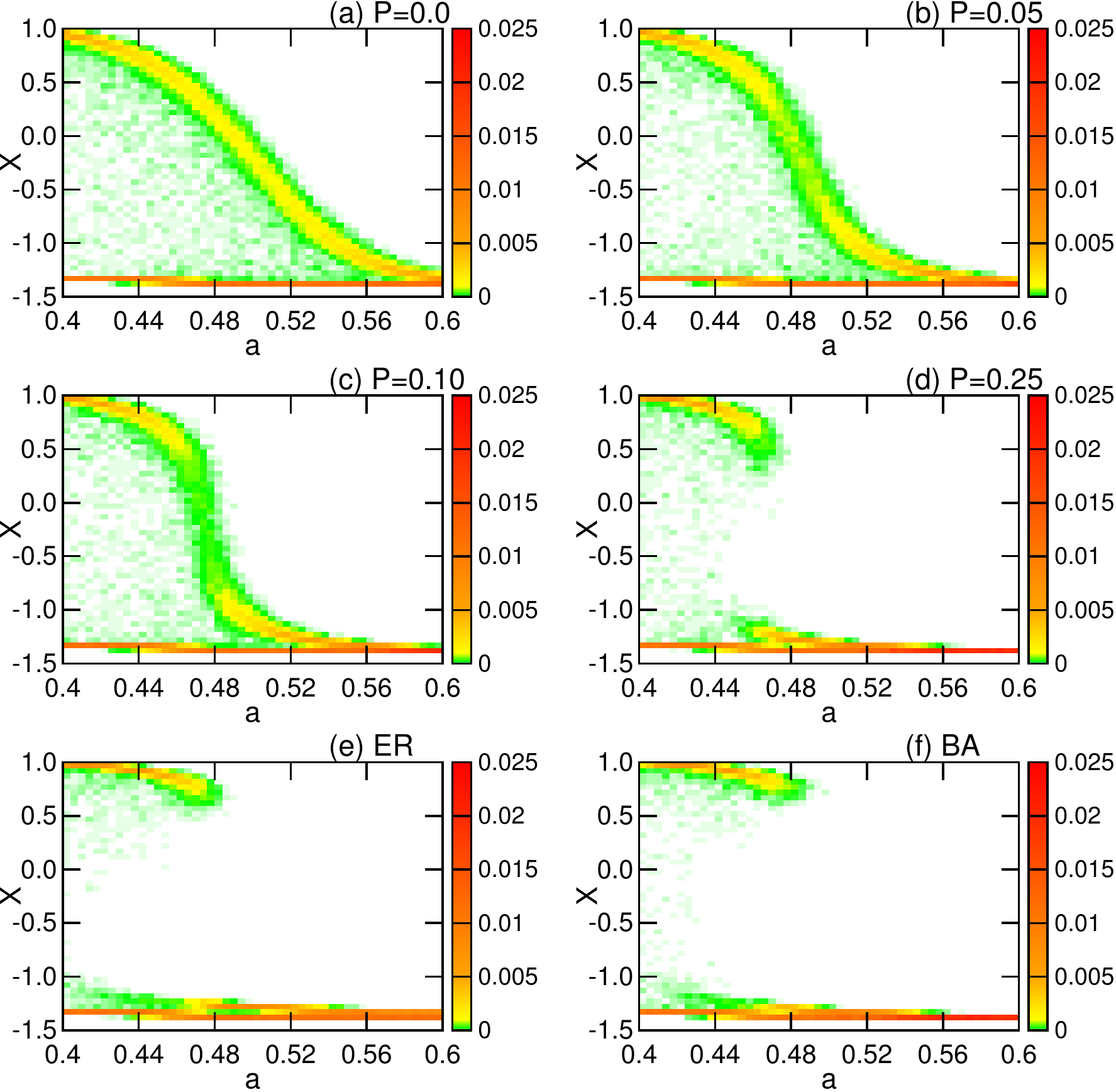}}
\caption{(color online). Stationary state distributions of the networks of the minimal systems. The Watts-Strogatz networks with (a) $P=0$, (b) $P=0.05$, (c) $P=0.1$, and (d) $P=0.25$. (e) The Erd\H{o}s-R\'{e}yni network. (f) The Barab\'asi-Albert network. For each panel, we generate $5 \times10^4$ stationary states with the stress $a$ and the initial condition $X_0$ randomly selected from uniform distributions $[0.4,0.6]$ and $[-1.6,1.6]$, respectively. (a)-(c) The upper attractors exist after the thresholds but the density of the stationary states having $-1.0<X<0.2$ decreases as $P$ increases in the regular networks. (d)-(f) The upper attractors disappear at the thresholds in the disordered networks.} \label{fig:3}
\end{figure}

The above results imply that the thresholds in the disordered networks are tipping points whereas the ones in the regular networks are not. To clarify the nature of the thresholds, we estimate the stationary state distribution of each network which manifests the network's attractor. We generate a large number of stationary states of the networks with the stress $a$ and the initial state $X_0$ randomly drawn from uniform distributions $[0.4,0.6]$ and $[-1.6,1.6]$, respectively. The state $X_0$ is assigned to all the nodes initially. The probability distribution of the generated stationary states in each network indicates that the attractors of the regular networks (Figs.~\ref{fig:3}(a)-\ref{fig:3}(c)) are fundamentally different from those of the disordered networks (Figs.~\ref{fig:3}(d)-\ref{fig:3}(f)). In the regular networks, the upper branch attractors continue to exist after the thresholds. But the density of stationary states having the values from $-1.0$ to $0.2$ becomes lower as $P$ increases in the WS networks (Figs.~\ref{fig:3}(b) and \ref{fig:3}(c)). In the disordered networks, the upper attractors suddenly disappear at the thresholds (Figs.~\ref{fig:3}(d)-(f)). This shows that the abrupt state changes in the disordered networks are also transitions between distinct attractors.

\begin{figure}[!ht]
\centerline{\includegraphics[width=0.99\linewidth,angle=0]{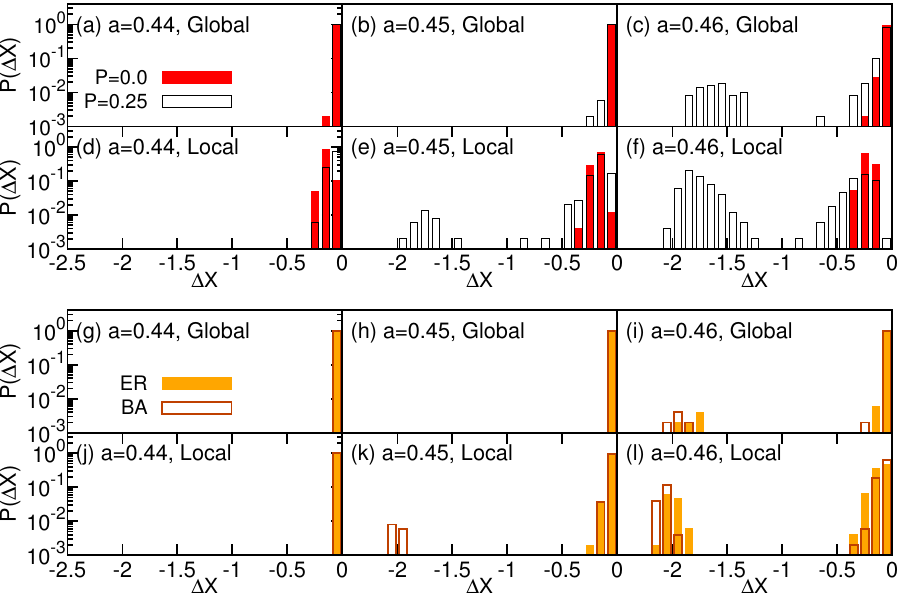}}
\caption{(color online). Distributions of state changes $\Delta X$ in the networks of the minimal systems after perturbations. (a)-(c) Global perturbations ($S_G=-0.1$) were given to all the nodes in the Watts-Strogatz (WS) networks with $P=0$ and $P=0.25$. (d)-(f) Local perturbations ($S_L=-1.0$) were given to randomly selected $10\%$ of the nodes in the WS networks with $P=0$ and $P=0.25$. (g)-(i) Global perturbations ($S_G=-0.1$) were given to all the nodes in the Erd\H{o}s-R\'{e}yni (ER) network and the Barab\'asi-Albert (BA) network. (j)-(l) Local perturbations ($S_L=-1.0$) were given to randomly selected $10\%$ of the nodes in the ER and the BA networks.} \label{fig:4}
\end{figure}

Another characteristic of tipping points is vulnerability to perturbations as depicted in Fig.~\ref{fig:1}(b). Small state changes can trigger to move systems to a contrasting state near catastrophic thresholds. Can we expect the same vulnerability in the disordered networks? To answer this, we perform perturbation experiments numerically. We obtain stationary states $X_{B}$ of a network for given stress $a$ and the initial condition $X_0$ randomly chosen from a uniform distribution $[1.0,1.6]$ and assigned equally to all the nodes. If $X_{B}>0$ (i.e., the network is active), then we give a small shock $S$ to the network and obtain the stationary state $X_{A}$ after the perturbation. We consider two types of perturbations. For the global perturbation, we give the same shock $S_G$ to all the nodes whereas we give the shock $S_L$ to randomly chosen $10\%$ of the nodes for the local perturbation. Figure~\ref{fig:4} represents the distributions of state changes  $\Delta X=X_{A}-X_{B}$ in the networks after the perturbations for the given stress $a$ and the perturbations. Regardless of the perturbation types, we found a consistent pattern. The regular networks are resilient against the perturbations in all the considered cases. The disordered networks become more likely to shift to different states after the perturbations as the stress approaches their thresholds.

\begin{figure}[!tpb]
\centerline{\includegraphics[width=0.98\linewidth,angle=0]{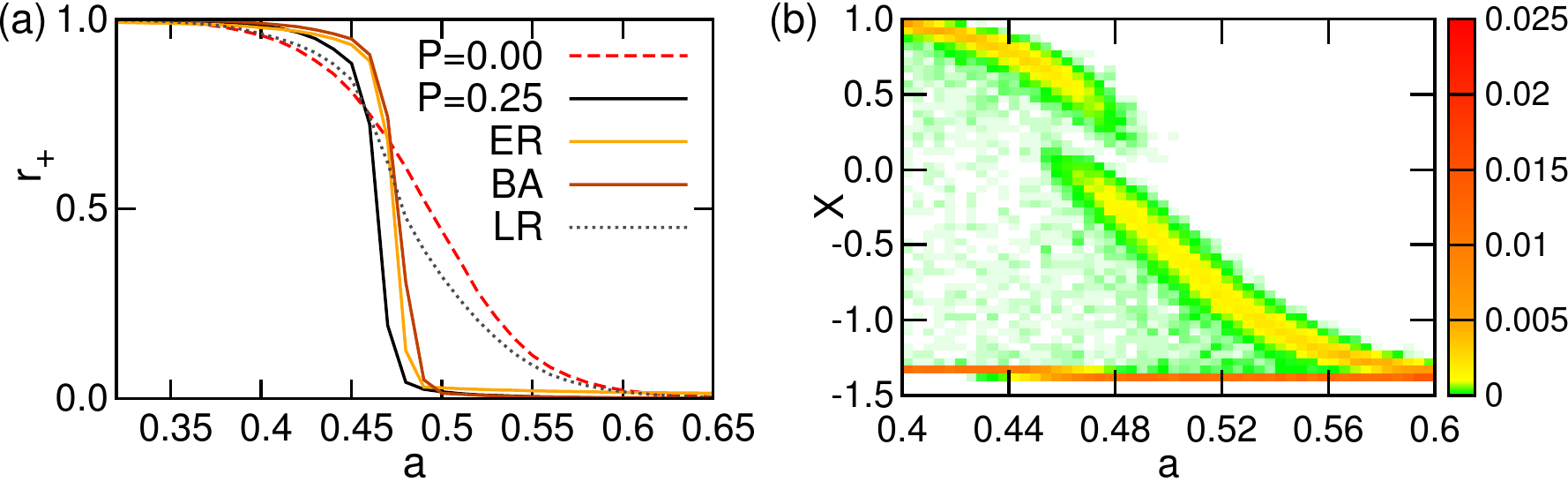}}
\caption{(color online). Suppressed collapse of individual nodes in the networks of the minimal systems. (a) The ratio of active nodes $r_+$ as a function of the stress $a$ shows the synchronized collapses of the nodes in the Watts-Strogatz (WS) network with $P=0.25$, the  Erd\H{o}s-R\'{e}yni network, and the Barab\'asi-Albert network but the gradual collapses of the nodes in the WS network with $P=0$ and the locally rewired (LR) networks. To generate the LR network, we divide a WS network into four equivalent sub-lattice before link rewiring and rewire only the links between the nodes in one of the sub-lattice. Thus the same number of links is rewired in the LR network as in the WS network with $P=0.25$. Each line is averaged over 50 realizations. (b) The stationary state distribution of the LR network shows a small gap indicating the collapse of the rewired sub-lattice in the upper attractor.} \label{fig:5}
\end{figure}

Our findings showed that the regular and the disordered networks show qualitatively different responses to environmental stress. Where does the difference come from? The coupling term of Eq.~(\ref{eq:2}) can bring a node the subsidy of activity from its neighboring nodes. This subsidy can suppress the collapse of individual nodes~\cite{Brummitt2015}. In the regular networks, the nodes can get only limited subsidy since the number of their neighboring nodes is few. However, the disordered networks have small-world topology~\cite{Watts1998} where the nodes can get network-wide subsidy because the average distance between the nodes is substantially short. Hence most nodes can suppress their collapses but eventually all the nodes collapse together. To validate such a trade-off between node and network resilience, we check the ratio of active nodes $r_+$ to the entire population as a function of the stress $a$ under the deteriorating scenario. As depicted in Fig.~\ref{fig:5}(a), the nodes in the regular networks started to crumble gradually at the lower thresholds whereas those in the disordered networks collapsed simultaneously at the higher thresholds. To check the effect of reduced distance, we divide a WS network into four equivalent sublattices before random link rewiring and rewire only the links between the nodes in one of the sublattices. This locally rewired (LR) network has the same number of rewired links as the original WS network with $P=0.25$ has, yet the average distance of the former is longer than that of the latter since the locally rewired links cannot provide long-range short-cuts. The nodes in the LR network gradually collapse (Fig.~\ref{fig:5}(a)). We generate stationary states of the LR network to obtain its attractor with the stress $a$ and the initial state $X_0$ randomly drawn from uniform distributions $[0.4,0.6]$ and $[-1.6,1.6]$, respectively. The upper attractors of the LR network has a small gap indicating the collapse of the rewired sub-lattice (Fig.~\ref{fig:5}(b)). Note that suppressing the onset of transitions also leads to other abrupt transitions in networks such as explosive percolation~\cite{Dsouza2015} and explosive synchronization~\cite{Zhang2015}.

Although we provide results only for the case of $D=1.0$ and $B=[0,3]$, and we consider the cusp catastrophe system as the dynamical model of the nodes here, our results are more general. To support the generality, we consider three additional cases. First we consider the same populations on the model networks but with the higher coupling strength $D=1.2$ (See Appendix B). Second we consider the mixed minimal systems whose growth rate $b$ is randomly drawn from a uniform distribution $B=[-1.5,1.5]$ (see Appendix C) rather than $B=[0,3]$ of the above analysis. As shown in Fig.~\ref{fig:1}, the minimal system responds abruptly when $b>0$ but gradually when $b<0$. Hence the half of the nodes responds gradually in these mixed populations when they are isolated. Third we consider the grazing system~\cite{NoyMeir1975,May1977}, a paradigmatic model for logistically growing vegetation under grazing stress (see Appendix D). The state of node $i$ in a network of $N$ grazing systems is evolved by
\begin{equation} \label{eq:3}
 \frac{dx_i}{dt} = x_i(1-\frac{x_i}{b_i})-\frac{ax_i^2}{x_i^2+1} + \frac{D}{k_i}\sum_{j=1}^{N} A_{ij}(x_j-x_i),
\end{equation}
where $x_i(\geq0)$ is the biomass density of node $i$, $a$ is the grazing stress, $b_i$ is the carrying capacity of node $i$ randomly drawn from a uniform distribution $B=[5.0,10.0]$, $D$ is the coupling strength ($D=0.3$). Hence the grazing systems with ecological implications are more realistic than the minimal systems. We consider these three additional cases on networks with $N=10^4$ and $\langle k\rangle=4$. As depicted in Appendices B, C, and D, we obtain qualitatively the same results demonstrating that, regardless of the some changes in the coupling strength, the heterogeneity, or the dynamical model of the nodes, the topological disorder of the networks can make their response to environmental stress more abrupt and lead to tipping points by inducing small-world topology in the networks. 

Last one may wonder the impact of the lattice dimension on the response of regular networks as we considered only the case of two-dimensional lattice. We can infer the impact of lattice dimension from our findings. The average distance, $\langle d \rangle_{r}$, among the nodes in regular lattices is given by $\langle d \rangle_{r} \sim N^{1/\alpha}$, where $N$ is the number of the nodes in the lattice and $\alpha$ is the lattice dimension~\cite{Barabasi2016book}. Hence, as $\alpha$ increases, the average distance decreases. Therefore, it is expected that high-dimensional regular networks can respond more abruptly to environmental stress than low-dimensional regular networks.

\section{Discussion and conclusions}
In conclusion, using a simple model, we have investigated the role of network topology in the response of networks to environmental stress, which has profound impacts on network resilience. We showed that network topology can determine how networks under environmental stress collapse at thresholds by regulating the trade-off between node resilience and network resilience to such stress. In the disordered networks, when the environments deteriorate, the nodes can persist in functioning owing to the networkwide subsidy caused by the small-world topology of the networks but eventually the whole networks abruptly collapse at tipping points in return. In the regular networks, some nodes fail in response to small environmental stress without the networkwide support because the networks are not small-world. Hence the regular networks gradually crumble as the stress increases. These results reveal the important role of small-world topology in network resilience to environmental stress. Note that network resilience to internal stress is mainly influenced by degree distributions or interdependency between networks~\cite{Albert2000,Cohen2000,Callaway2000,Cohen2001,Vazquez2003,Valente2004,Gallos2005,Buldyrev2010,Parshani2010,Gao2011,Gao2011a,Bashan2013}. Such a difference between the response to environmental stress and the response to internal stress indicate that a network may respond gradually to one type of stress but it can respond abruptly to the other stress. For example, a BA network is known to be robust to the random failures of its nodes owing to its broad degree distribution whereas it can collapse abruptly as a result of small changes in environmental stress because of its small-world topology. Therefore it is necessary to take into account the impact of both internal and environmental stress for deeper understanding of network resilience.

Our findings indicate that some real networks can be highly resilient against environmental stress to a threshold yet extremely vulnerable to the stress above the threshold because they have small-world topology. In such cases, the resilience to a certain level of the stress cannot guarantee the resilience to the slightly larger stress. Instead, network can collapse abruptly at the slightly larger stress. Early-warning indicators for such vulnerability need to be developed because it is difficult to reverse once the networks collapsed. Critical slowing down (i.e., slower recovery from perturbations in vicinity of bifurcation points) and its related indicators have been suggested as early-warning indicators for catastrophic shifts in some ecosystems, microorganism populations, and mutualistic networks~\cite{Scheffer2009a,Dai2012,Drake2010,Carpenter2011,Scheffer2012,Dakos2014}. This prompts us to check whether a similar phenomenon can arise in our model to foresee the sudden collapse of networks as a future work. Last our results suggest that whether growing environmental stress is a crucial threat or not can depend on other topological properties of networks. In fact, real networks have not only small-world topology but also other common properties such as degree-correlations, community structure, and broad degree distributions, to name a few~\cite{Barabasi2016book}. We wish to study how such characteristics determine the response of real networks to environmental stress to design or reshape network topology for more resilient networks as future works. 

\appendix  




\section{The response of regular networks with different parameters}

\begin{figure}[!tpb]
\centerline{\includegraphics[width=0.99\linewidth,angle=0]{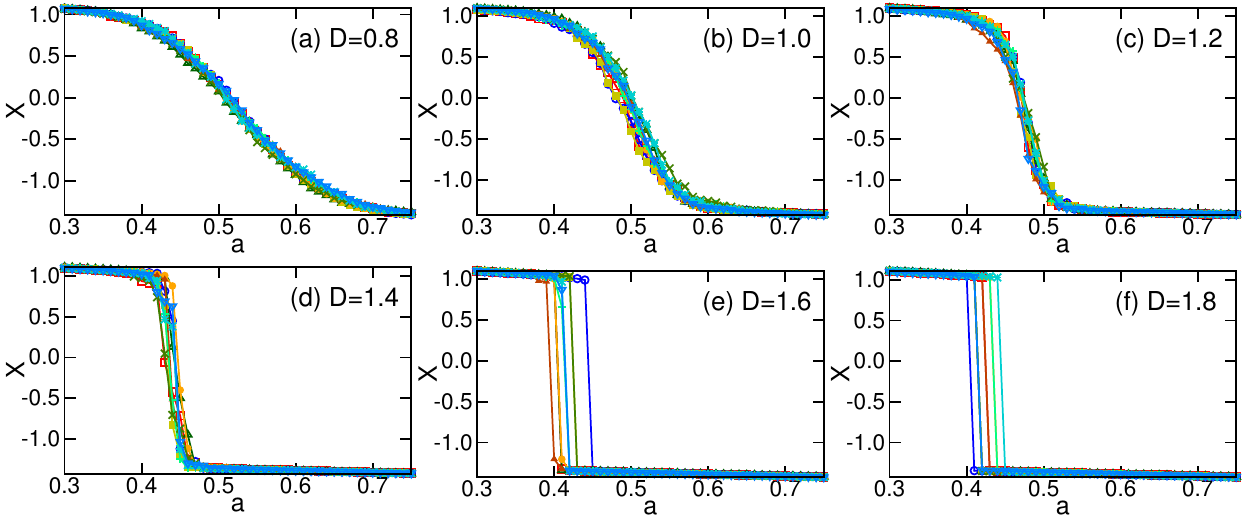}}
\caption{(color online). State diagrams $X(a)$ of the regular networks of the minimal systems. All networks are generated from the two-dimensional Watts-Strogatz model with $P=0.0$. As the coupling strength $D$ increases, the response of network states to environmental stress $a$ becomes abrupt. The parameter values are $N=10^4$, $B=[0,3]$, and $\langle k \rangle=4$. (a) $D=0.8$. (b) $D=1.0$. (c) $D=1.2$. (d) $D=1.4$. (e) $D=1.6$. (f) $D=1.8$. Each curve shows a single realization. Ten realizations are represented for each panel.} \label{fig:r1}
\end{figure}

\begin{figure}[!tpb]
\centerline{\includegraphics[width=0.99\linewidth,angle=0]{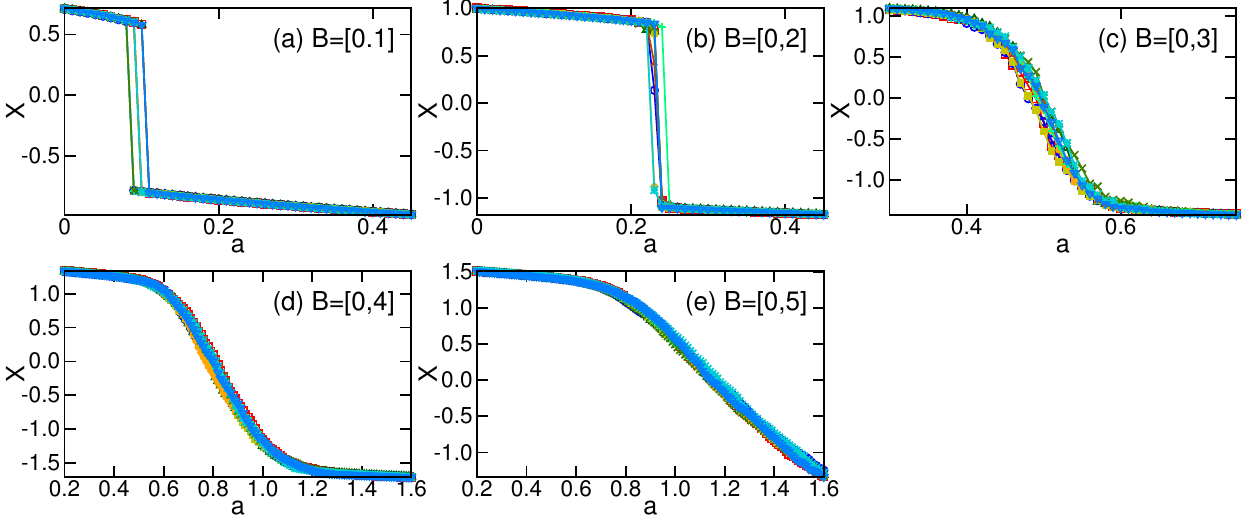}}
\caption{(color online). State diagrams $X(a)$ of the regular networks of the minimal systems. All networks are generated from the two-dimensional Watts-Strogatz model with $P=0.0$. As the width of the interval $B$ decreases, the response of network states to environmental stress $a$ becomes abrupt. The parameter values are $N=10^4$, $D=1.0$, and $\langle k \rangle=4$. (a) $B=[0,1]$.  (b) $B=[0,2]$. (c) $B=[0,3]$. (d) $B=[0,4]$. (e) $B=[0,5]$. Each curve shows a single realization. Ten realizations are represented for each panel.} \label{fig:r2}
\end{figure}

\begin{figure}[!b]
\centerline{\includegraphics[width=0.99\linewidth,angle=0]{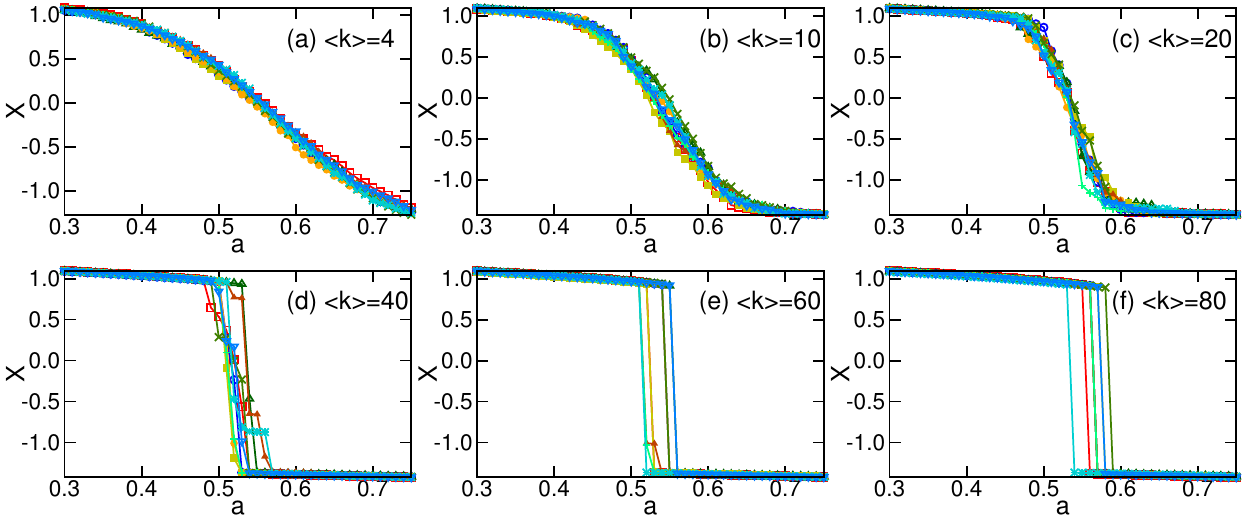}}
\caption{(color online). State diagrams $X(a)$ of the regular networks of the minimal systems. All networks are generated from the one-dimensional Watts-Strogatz model with $P=0.0$. Note that, unlike the other cases, we consider one-dimensional Watts-Strogatz model to change the average degree in the networks. As the average degree increases, the response of network states to environmental stress $a$ becomes abrupt. The parameter values are $N=10^4$, $B=[0,3]$, and $D=1.0$. (a) $\langle k \rangle=4$.  (b) $\langle k \rangle=10$. (c) $\langle k \rangle=20$. (d) $\langle k \rangle=40$. (e) $\langle k \rangle=60$. (f) $\langle k \rangle=80$. Each curve shows a single realization. Ten realizations are represented for each panel.} \label{fig:r3}
\end{figure}

We consider networks of the same minimal systems defined in the main text. Hence the state of node $i$ in a network of $N$ minimal systems is evolved by 
\begin{equation} \label{eq:S0}
 \frac{dx_i}{dt} = -a+b_ix_i-{x_i}^{3} + \frac{D}{k_i}\sum_{j=1}^N A_{ij}(x_j-x_i),
\end{equation}
where $x_i$ is the state of node $i$, $a$ is the environmental stress, $b_i$ is the growth rate of node $i$ drawn from a uniform distribution $B$, $D$ is the overall coupling strength among the nodes, $k_i$ is the degree of node $i$, and $A_{ij}$ is the element of the adjacency matrix of the network. The network's state $X$ is defined as $X=\frac{1}{N}\sum_i x_i$. In this section we focus on the response of regular networks to environmental stress. Regular networks can respond abruptly to environmental stress when the heterogeneity in the response of the individual components is weak (i.e., the $B$ is narrow) or the connectivity between the components are high (i.e., the coupling strength $D$ or the average degree $\langle k \rangle$ are high)~\cite{vanNes2005,Martin2015}. We consider a deteriorating scenario where the stationary value of $X(a)$ is obtained by increasing the stress $a$ with an increment $\delta a=0.01$ from $a=-0.8$ to $a=0.8$. We observed that the response of the networks generated by the Watts-Strogatz model with probability $P=0$ gets more abrupt when the coupling strength $D$ increases (Fig.~\ref{fig:r1}), the heterogeneity of the growth rates decreases (Fig.~\ref{fig:r2}), or the average degree increases (Fig.~\ref{fig:r3}). Note that we consider one-dimensional Watts-Strogatz model in Fig.~\ref{fig:r3} unlike in Figs.~\ref{fig:r1} and~\ref{fig:r2} since one-dimensional case is easy to control the average degree than two-dimensional case in the main text.


\section{Networks of the minimal systems with different coupling strength}

\begin{figure}[!tpb]
\centerline{\includegraphics[width=0.99\linewidth,angle=0]{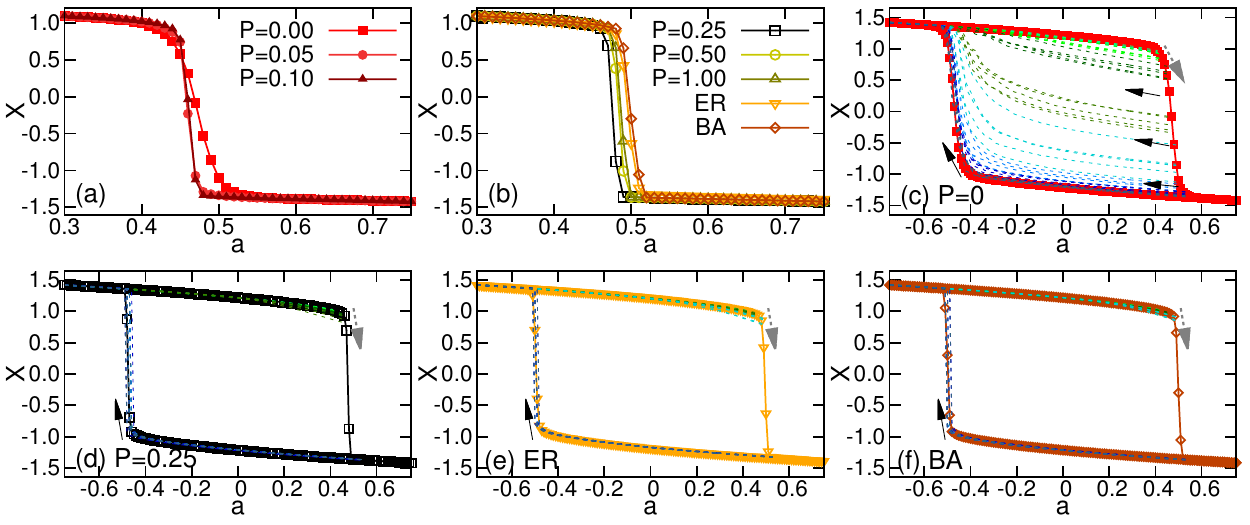}}
\caption{(color online). State diagrams $X(a)$ of the networks of the minimal systems with $D=1.2$. (a) The Watts-Strogatz (WS) network with $P=0.0$ responds gradually to the increasing stress $a$. But, as $P$ increases, the response becomes more abrupt. (b) The WS networks with $P\geq 0.25$, the Erd\H{o}s-R\'{e}yni (ER), and the Barab\'asi-Albert (BA) respond abruptly to the increasing stress. Increasing and then decreasing the stress, we observed hysteresis patterns in (c) the WS network with $P=0$, (d) the WS network with $P=0.25$, (e) the ER network, and (f) the BA network. Each line with symbols is averaged over 50 realizations. The dotted lines in (c)-(f) show that the hysteresis can be mitigated in the WS network with $P=0$ but not in the WS network with $P=0.25$, the ER network, and the BA network. Each dotted line shows a single realization obtained by increasing and then decreasing the stress $a$ when it reaches $a=0.41$, $0.43$, $0.45$, $0.47$, $0.49$, $0.51$, $0.53$, or $0.55$.} \label{fig:SM01}
\end{figure}

\begin{figure}[!tpb]
\centerline{\includegraphics[width=0.99\linewidth,angle=0]{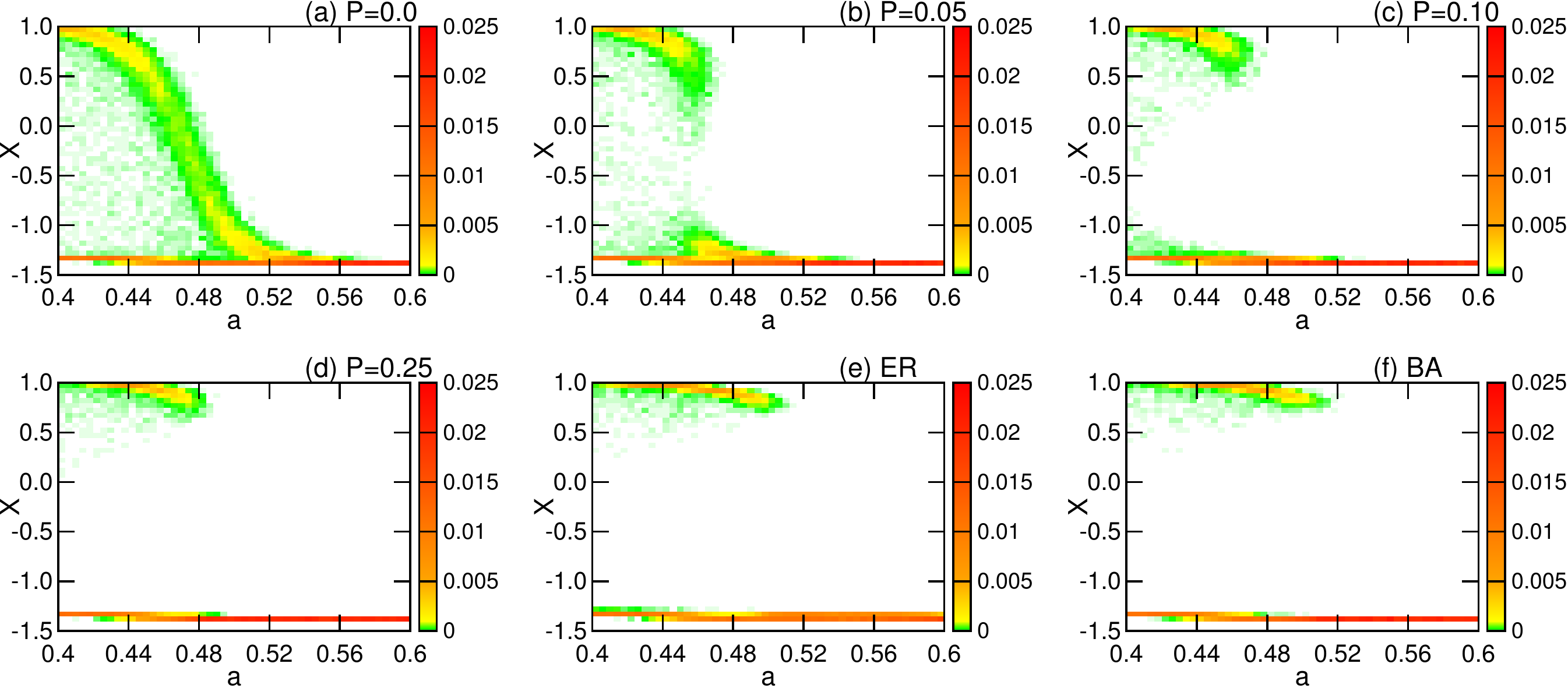}}
\caption{(color online). Stationary state distributions of the networks of the minimal systems with $D=1.2$. The Watts-Strogatz networks with (a) $P=0$, (b) $P=0.05$, (c) $P=0.1$, and (d) $P=0.25$. (e) The Erd\H{o}s-R\'{e}yni network. (f) The Barab\'asi-Albert network. For each panel, we generate $5 \times10^4$ stationary states with the stress $a$ and the initial condition $X_0$ randomly selected from uniform distributions $[0.4,0.6]$ and $[-1.6,1.6]$, respectively. (a)-(b) The upper attractors exist after the thresholds but the density of the stationary states having $-1.0<X<0.2$ decreases as $P$ increases in the regular networks. (c)-(f) The upper attractors disappear at the thresholds in the disordered networks.} \label{fig:SM02}
\end{figure}

\begin{figure}[!b]
\centerline{\includegraphics[width=0.99\linewidth,angle=0]{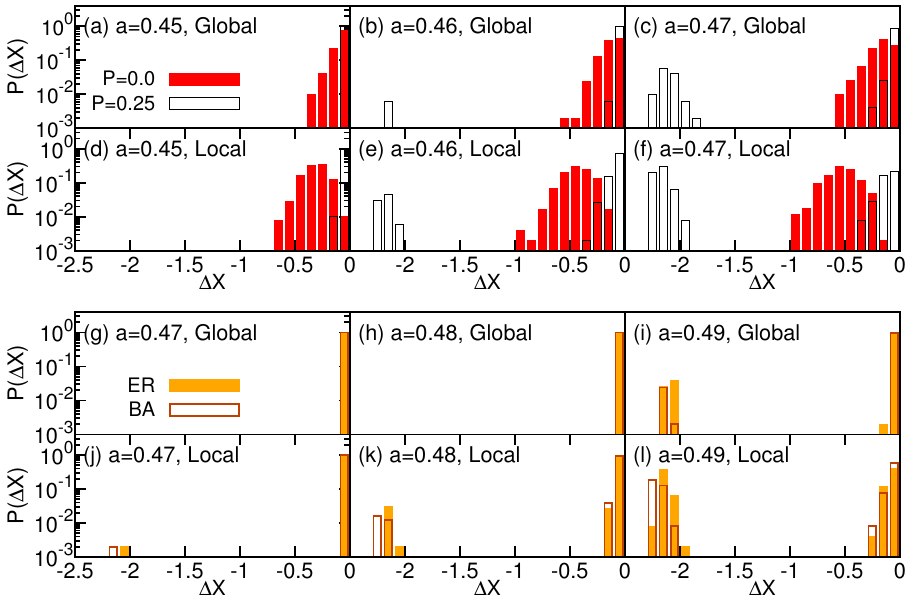}}
\caption{(color online). Distributions of state changes $\Delta X=X_A-X_B$ in the networks of the minimal systems with $D=1.2$ after perturbations. We obtain stationary states $X_{B}$ of a network for given stress $a$ and the initial condition $X_0$ randomly chosen from a uniform distribution $[1.0,1.6]$ and assigned equally to all the nodes. If $X_{B}>0$ (i.e., the network is active), we give a small shock $S$ to the network and obtain the stationary state $X_{A}$ after the perturbation. (a)-(c) Global perturbations ($S_G=-0.1$) were given to all the nodes in the Watts-Strogatz (WS) networks with $P=0$ and $P=0.25$. (d)-(f) Local perturbations ($S_L=-1.0$) were given to randomly selected $10\%$ of the nodes in the WS networks with $P=0$ and $P=0.25$. (g)-(i) Global perturbations ($S_G=-0.1$) were given to all the nodes in the Erd\H{o}s-R\'{e}yni (ER) network and the Barab\'asi-Albert (BA) network. (j)-(l) Local perturbations ($S_L=-1.0$) were given to randomly selected $10\%$ of the nodes in the ER and the BA networks.} \label{fig:SM03}
\end{figure}

\begin{figure}[!tpb]
\centerline{\includegraphics[width=0.99\linewidth,angle=0]{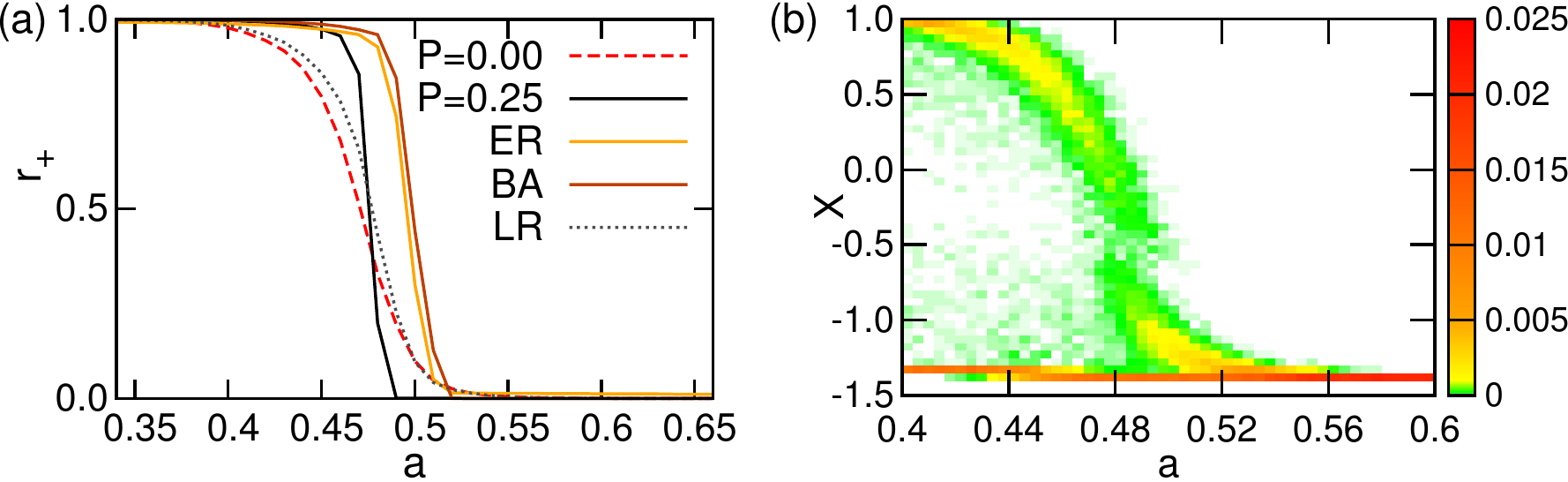}}
\caption{(color online). Suppressed collapse of individual nodes in the networks of the minimal systems with $D=1.2$. (a) The ratio of active nodes $r_+$ as a function of the stress $a$ shows the synchronized collapses of the nodes in the Watts-Strogatz (WS) network with $P=0.25$, the  Erd\H{o}s-R\'{e}yni network, and the Barab\'asi-Albert network but the gradual collapses of the nodes in the WS network with $P=0$ and the locally rewired (LR) networks. To generate the LR network, we divide a WS network into four equivalent sub-lattice before link rewiring and rewire only the links between the nodes in one of the sub-lattice. Thus the same number of links is rewired in the LR network as in the WS network with $P=0.25$. Each line is averaged over 50 realizations. (b) The stationary state distribution of the LR network shows a small gap indicating the collapse of the rewired region in the upper attractor. We generate $5 \times10^4$ stationary states with the stress $a$ and the initial condition $X_0$ randomly selected from uniform distributions $[0.4,0.6]$ and $[-1.6,1.6]$, respectively.} \label{fig:SM04}
\end{figure}

We consider networks of the same minimal systems defined in the main text. Hence the state of node $i$ in a network of $N$ minimal systems is evolved by 
\begin{equation} \label{eq:S1}
 \frac{dx_i}{dt} = -a+b_ix_i-{x_i}^{3} + \frac{D}{k_i}\sum_{j=1}^N A_{ij}(x_j-x_i),
\end{equation}
where $x_i$ is the state of node $i$, $a$ is the environmental stress, $b_i$ is the growth rate of node $i$ drawn from a uniform distribution $B$, $D$ is the overall coupling strength among the nodes, $k_i$ is the degree of node $i$, and $A_{ij}$ is the element of the adjacency matrix of the network. The network's state $X$ is defined as $X=\frac{1}{N}\sum_i x_i$. We consider the networks with $N=10^4$, $B=[0,3]$, and $\langle k\rangle=4$ same as the ones in the main text. The only difference is that the coupling strength is changed to $D=1.2$ whereas we took $D=1.0$ in the main text. In Fig.~\ref{fig:SM01} we consider a deteriorating (ameliorating) scenario where the stationary value of $X(a)$ is obtained by increasing the stress $a$ with an increment (decrement) $\delta a=0.01$ from $a=-0.8$ ($a=0.8$) to $a=0.8$ ($a=-0.8$). The information on the other numeric experiments is given in the caption of each figure. Figure~\ref{fig:SM02} represents the stationary state distributions of the networks of the minimal systems with $D=1.2$. Figure~\ref{fig:SM03} depicts the distributions of state changes $\Delta X=X_A-X_B$ in the networks of the minimal systems with $D=1.2$ after perturbations. Figure~\ref{fig:SM04} shows the suppressed collapse of individual nodes in the networks of the minimal systems with $D=1.2$.

\section{Networks of the mixed minimal systems}
\begin{figure}[!tpb]
\centerline{\includegraphics[width=0.99\linewidth,angle=0]{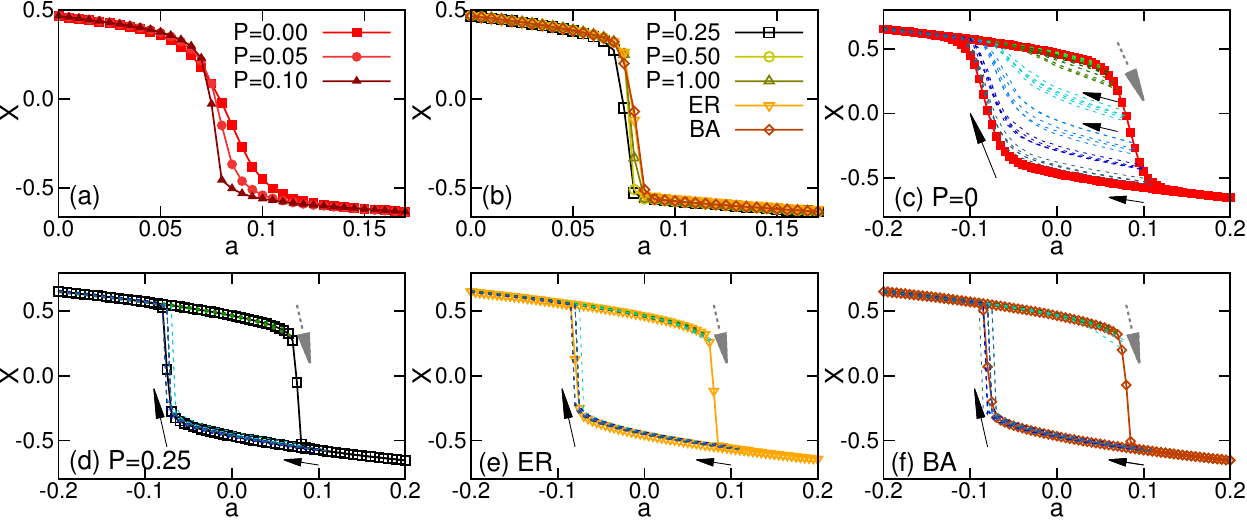}}
\caption{(color online). State diagrams $X(a)$ of the networks of the mixed minimal systems with $B=[-1.5,1.5]$. (a) The Watts-Strogatz (WS) networks with $P\leq0.1$ respond gradually to the increasing stress $a$. (b) The WS networks with $P\geq 0.25$, the Erd\H{o}s-R\'{e}yni (ER), and the Barab\'asi-Albert (BA) respond abruptly to the increasing stress. Increasing and then decreasing the stress, we observed hysteresis patterns in (c) the WS network with $P=0$, (d) the WS network with $P=0.25$, (e) the ER network, and (f) the BA network. Each line with symbols is averaged over 50 realizations. The dotted lines in (c)-(f) show that the hysteresis can be mitigated in the WS network with $P=0$ but not in the WS network with $P=0.25$, the ER network, and the BA network. Each dotted line shows a single realization obtained by increasing and then decreasing the stress $a$ when it reaches $a=0.04$, $0.05$, $0.06$, $0.07$, $0.08$, $0.09$, $0.1$, or $0.11$.} \label{fig:SM1}
\end{figure}

\begin{figure}[!b]
\centerline{\includegraphics[width=0.99\linewidth,angle=0]{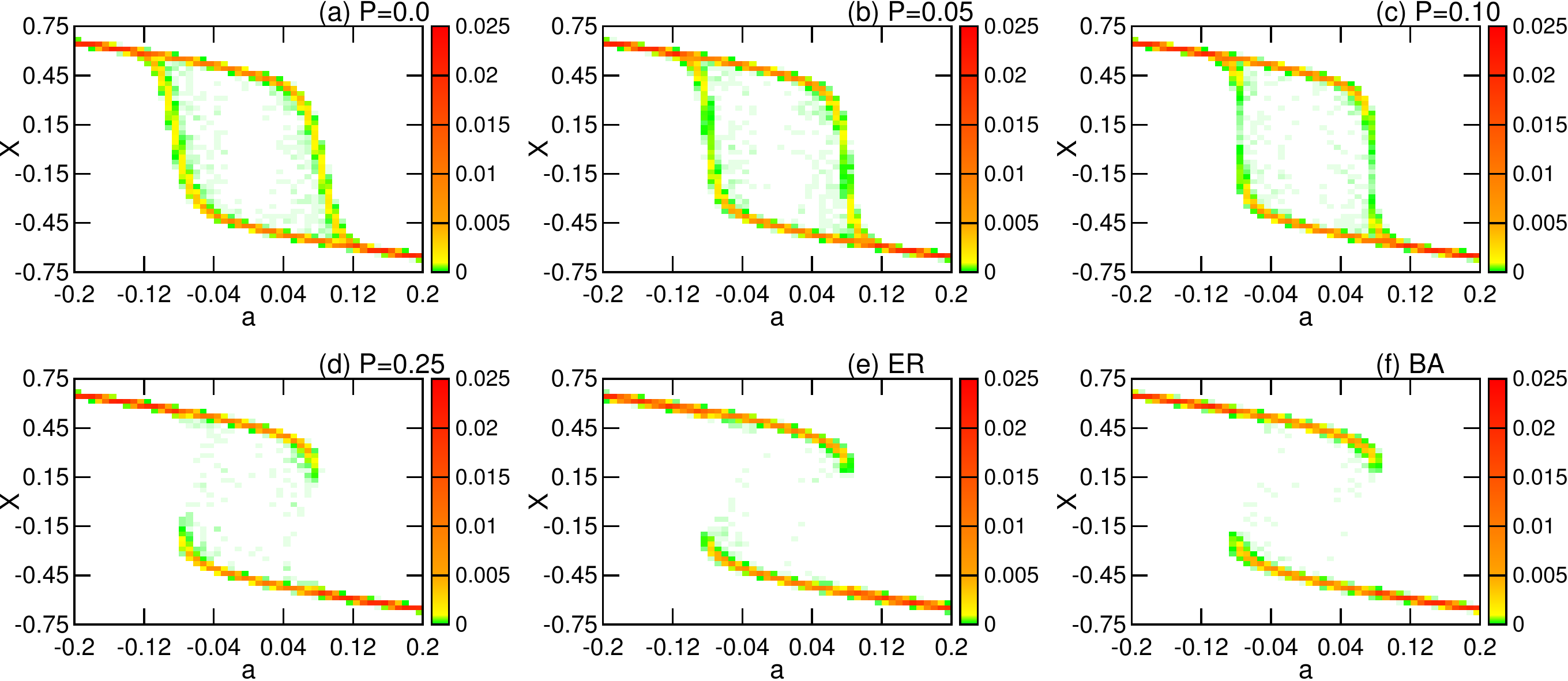}}
\caption{(color online). Stationary state distributions of the networks of the mixed minimal systems with $B=[-1.5,1.5]$. The Watts-Strogatz networks with (a) $P=0$, (b) $P=0.05$, (c) $P=0.1$, and (d) $P=0.25$. (e) The Erd\H{o}s-R\'{e}yni network. (f) The Barab\'asi-Albert network. For each panel, we generate $5 \times10^4$ stationary states with the stress $a$ and the initial condition $X_0$ randomly selected from uniform distributions $[-0.2,0.2]$ and $[-1.6,1.6]$, respectively. (a)-(c) The upper attractors exist after the thresholds but the density of the stationary states having $-0.3<X<0.3$ decreases as $P$ increases in the regular networks. (d)-(f) The upper attractors disappear at the thresholds in the disordered networks.} \label{fig:SM2}
\end{figure}

\begin{figure}[!tpb]
\centerline{\includegraphics[width=0.99\linewidth,angle=0]{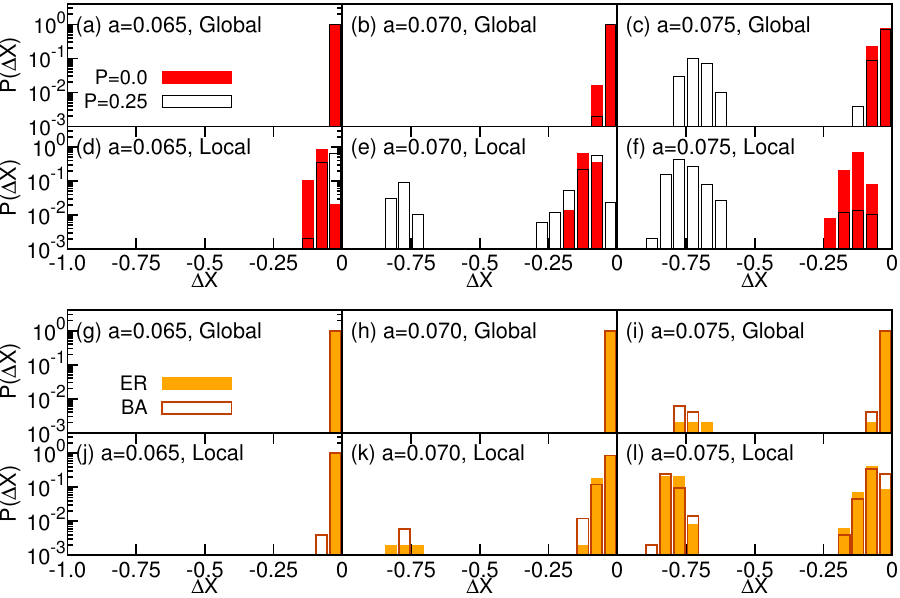}}
\caption{(color online). Distributions of state changes $\Delta X=X_A-X_B$ in the networks of the mixed minimal systems with $B=[-1.5,1.5]$ after perturbations. We obtain stationary states $X_{B}$ of a network for given stress $a$ and the initial condition $X_0$ randomly chosen from a uniform distribution $[1.0,1.6]$ and assigned equally to all the nodes. If $X_{B}>0$ (i.e., the network is active), then we give a small shock $S$ to the network and obtain the stationary state $X_{A}$ after the perturbation. (a)-(c) Global perturbations ($S_G=-0.1$) were given to all the nodes in the Watts-Strogatz (WS) networks with $P=0$ and $P=0.25$. (d)-(f) Local perturbations ($S_L=-1.0$) were given to randomly selected $10\%$ of the nodes in the WS networks with $P=0$ and $P=0.25$. (g)-(i) Global perturbations ($S_G=-0.1$) were given to all the nodes in the Erd\H{o}s-R\'{e}yni (ER) network and the Barab\'asi-Albert (BA) network. (j)-(l) Local perturbations ($S_L=-1.0$) were given to randomly selected $10\%$ of the nodes in the ER and the BA networks.} \label{fig:SM3}
\end{figure}

\begin{figure}[!tpb]
\centerline{\includegraphics[width=0.99\linewidth,angle=0]{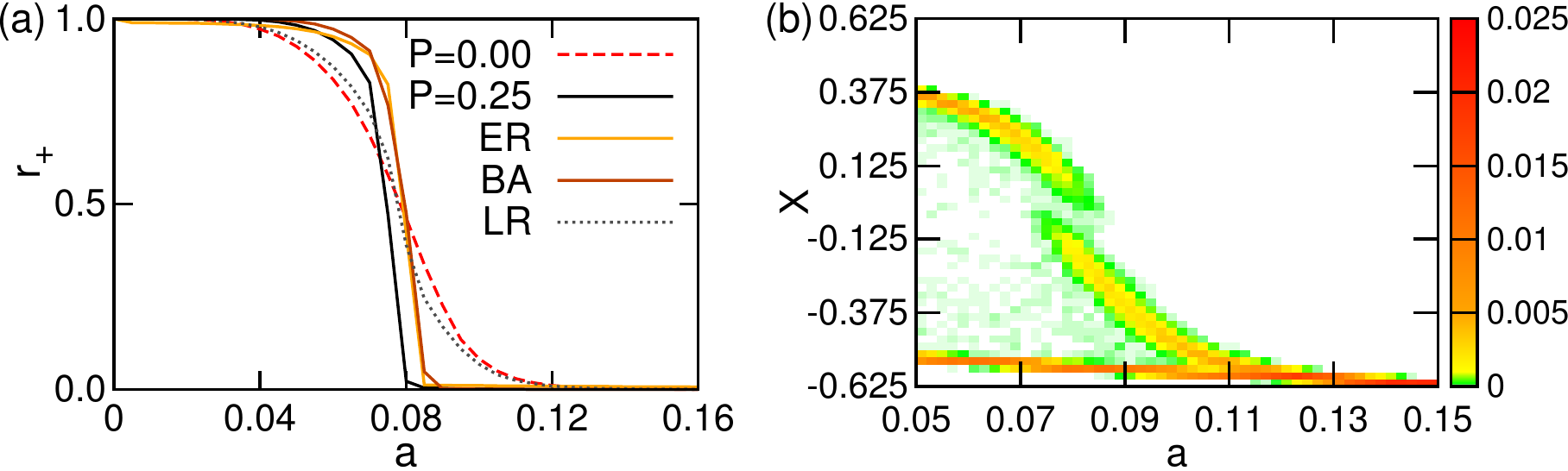}}
\caption{(color online). Suppressed collapse of individual nodes in the networks of the mixed minimal systems with $B=[-1.5,1.5]$. (a) The ratio of active nodes $r_+$ as a function of the stress $a$ shows the synchronized collapses of the nodes in the Watts-Strogatz (WS) network with $P=0.25$, the  Erd\H{o}s-R\'{e}yni network, and the Barab\'asi-Albert network but the gradual collapses of the nodes in the WS network with $P=0$ and the locally rewired (LR) networks. To generate the LR network, we divide a WS network into four equivalent sub-lattice before link rewiring and rewire only the links between the nodes in one of the sublattice. Thus the same number of links is rewired in the LR network as in the WS network with $P=0.25$. Each line is averaged over 50 realizations. (b) The stationary state distribution of the LR network shows a small gap indicating the collapse of the rewired region in the upper attractor. We generate $5 \times10^4$ stationary states with the stress $a$ and the initial condition $X_0$ randomly selected from uniform distributions $[0.05,0.15]$ and $[-1.6,1.6]$, respectively.} \label{fig:SM_MixSup}
\end{figure}

We consider networks of the same minimal systems defined in the main text. Hence the state of node $i$ in a network of $N$ minimal systems is evolved by 
\begin{equation} \label{eq:S2}
 \frac{dx_i}{dt} = -a+b_ix_i-{x_i}^{3} + \frac{D}{k_i}\sum_{j=1}^N A_{ij}(x_j-x_i),
\end{equation}
where $x_i$ is the state of node $i$, $a$ is the environmental stress, $b_i$ is the growth rate of node $i$ drawn from a uniform distribution $B$, $D$ is the overall coupling strength among the nodes, $k_i$ is the degree of node $i$, and $A_{ij}$ is the element of the adjacency matrix of the network. The network's state $X$ is defined as $X=\frac{1}{N}\sum_i x_i$. We consider the networks with $N=10^4$, $D=1.0$, and $\langle k\rangle=4$ same as the ones in the main text. The only difference is the uniform distribution $B$ for the growth rate of the nodes is changed to $B=[-1.5,1.5]$ whereas $B=[0,3]$ in the main text. This means that all nodes respond abruptly when isolated in the main text. On the other hand, here, the half of the nodes responds gradually and the other half responds abruptly when the nodes are isolated. In Fig.~\ref{fig:SM1} we consider a deteriorating (ameliorating) scenario where the stationary value of $X(a)$ is obtained by increasing the stress $a$ with an increment (decrement) $\delta a=0.005$ from $a=-0.2$ ($a=0.2$) to $a=0.2$ ($a=-0.2$). The information on the other numeric experiments is given in the caption of each figure. Figure~\ref{fig:SM2} represents the stationary state distributions of the networks of the mixed minimal systems with $B=[-1.5,1.5]$. Figure~\ref{fig:SM3} depicts the distributions of state changes $\Delta X=X_A-X_B$ in the networks of the mixed minimal systems with $B=[-1.5,1.5]$ after perturbations. Figure~\ref{fig:SM_MixSup} shows the suppressed collapse of individual nodes in the networks of the mixed minimal systems with $B=[-1.5,1.5]$.

\section{Networks of the grazing systems}
\begin{figure}[!h]
\centerline{\includegraphics[width=0.9\linewidth,angle=0]{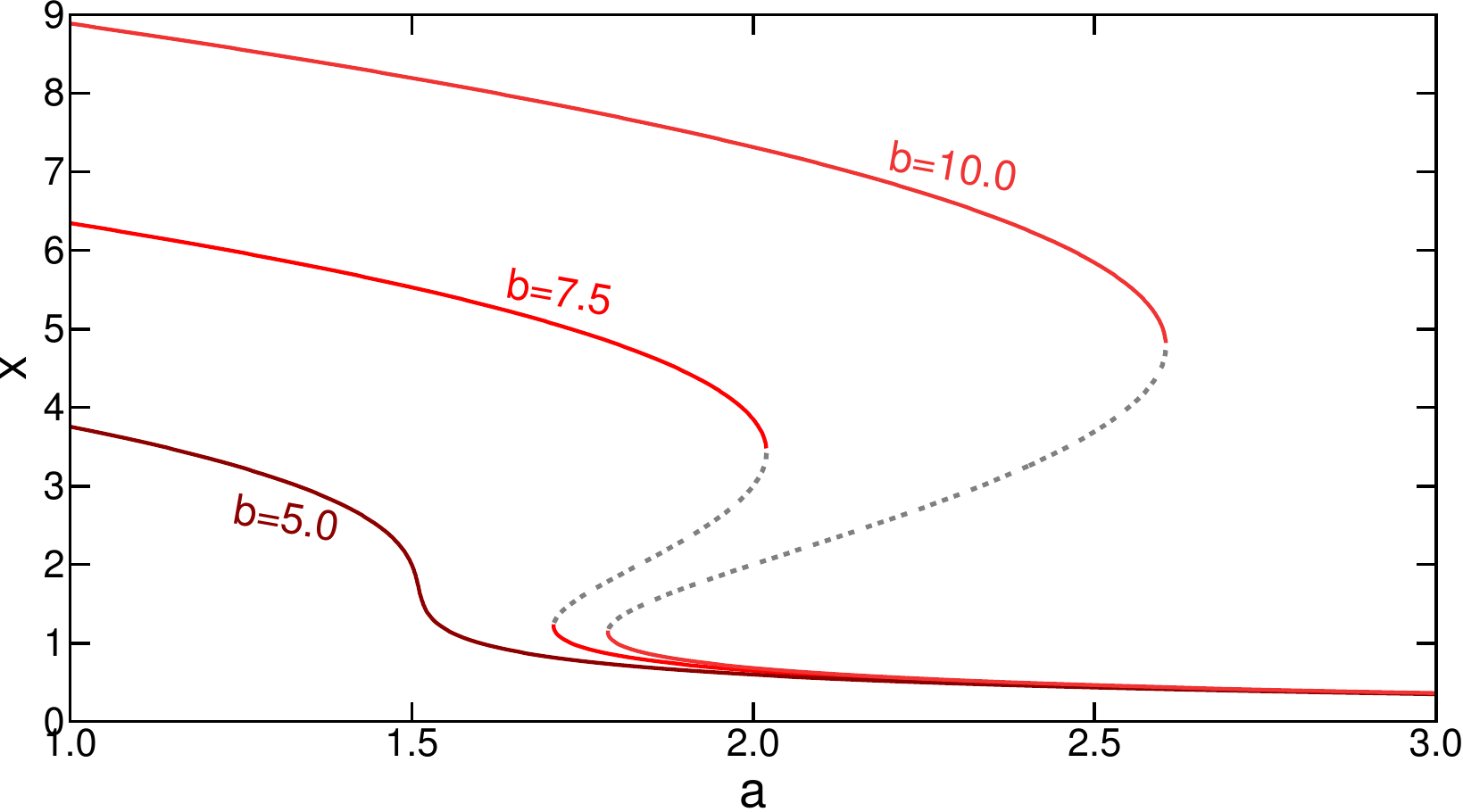}}
\caption{(color online). Bifurcation diagram of a grazing system whose dynamics is described by $dx/dt=x(1-x/b)-ax^2/(x^2+1)$. The red lines represent the stable states and the dotted gray line represents the unstable states.} \label{fig:SBDGr}
\end{figure}

\begin{figure}[!ht]
\centerline{\includegraphics[width=0.99\linewidth,angle=0]{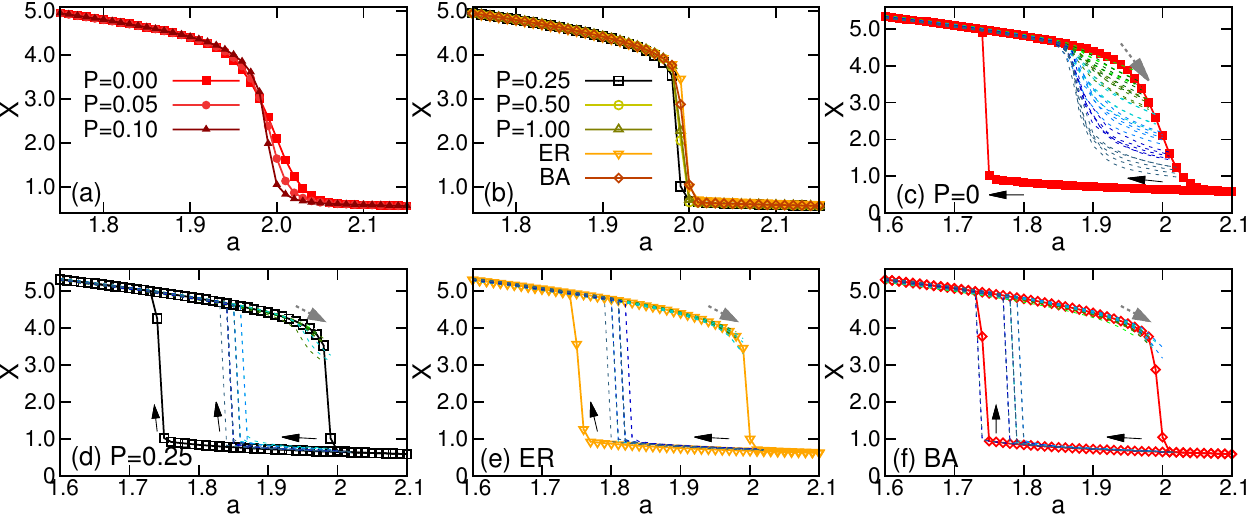}}
\caption{(color online). State diagrams $X(a)$ of the networks of the grazing systems. (a) The Watts-Strogatz (WS) networks with $P\leq0.1$ respond gradually to the increasing stress $a$. (b) The WS networks with $P\geq 0.25$, the Erd\H{o}s-R\'{e}yni (ER), and the Barab\'asi-Albert (BA) respond abruptly to the increasing stress. Increasing and then decreasing the stress, we observed hysteresis patterns in (c) the WS network with $P=0$, (d) the WS network with $P=0.25$, (e) the ER network, and (f) the BA network. Each line with symbols is averaged over 50 realizations. The dotted lines in (c)-(f) show that the hysteresis can be substantially mitigated in the WS network with $P=0$. But only weak mitigation is observed in the WS network with $P=0.25$, the ER network, and the BA network because some nodes with high growth rates (e.g., nodes with $b>9$) are still active after the thresholds and they can provide strong subsidy to other nodes as shown in Fig.~\ref{fig:SBDGr}. Each dotted line shows a single realization obtained by increasing and then decreasing the stress $a$ when it reaches $a=1.96$, $1.97$, $1.98$, $1.99$, $2.00$, $2.01$, or $2.02$.} \label{fig:SM4}
\end{figure}

\begin{figure}[!ht]
\centerline{\includegraphics[width=0.99\linewidth,angle=0]{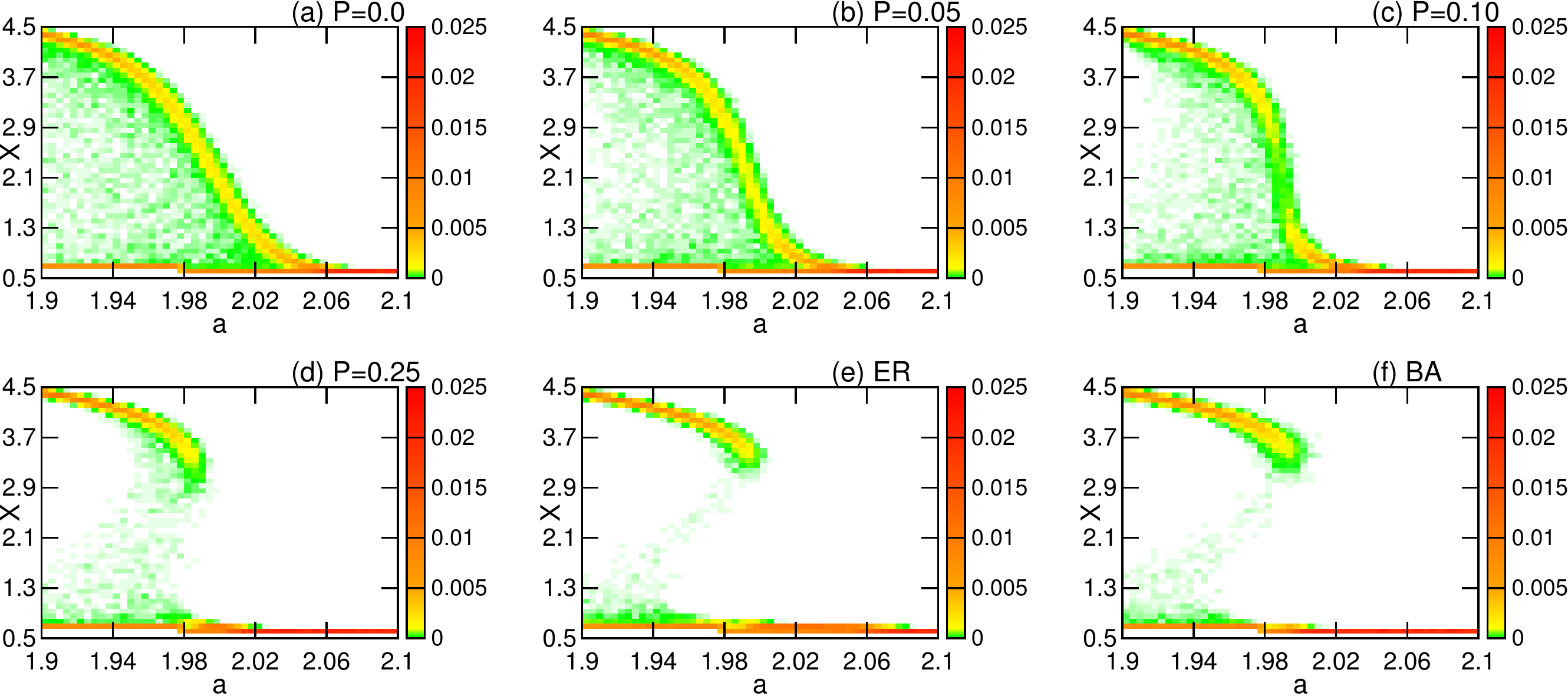}}
\caption{(color online). Stationary state distributions of the networks of the grazing systems. The Watts-Strogatz networks with (a) $P=0$, (b) $P=0.05$, (c) $P=0.1$, and (d) $P=0.25$. (e) The Erd\H{o}s-R\'{e}yni network. (f) The Barab\'asi-Albert network. For each panel, we generate $5 \times10^4$ stationary states with the stress $a$ and the initial condition $X_0$ randomly selected from uniform distributions $[1.9,2.1]$ and $[0.1,5.1]$, respectively. (a)-(c) The upper attractors exist after the thresholds but the density of the stationary states having $1.3<X<2.9$ decreases as $P$ increases in the regular networks. (d)-(f) The upper attractors disappear at the thresholds in the disordered networks.} \label{fig:SM5}
\end{figure}

\begin{figure}[!tpb]
\centerline{\includegraphics[width=0.99\linewidth,angle=0]{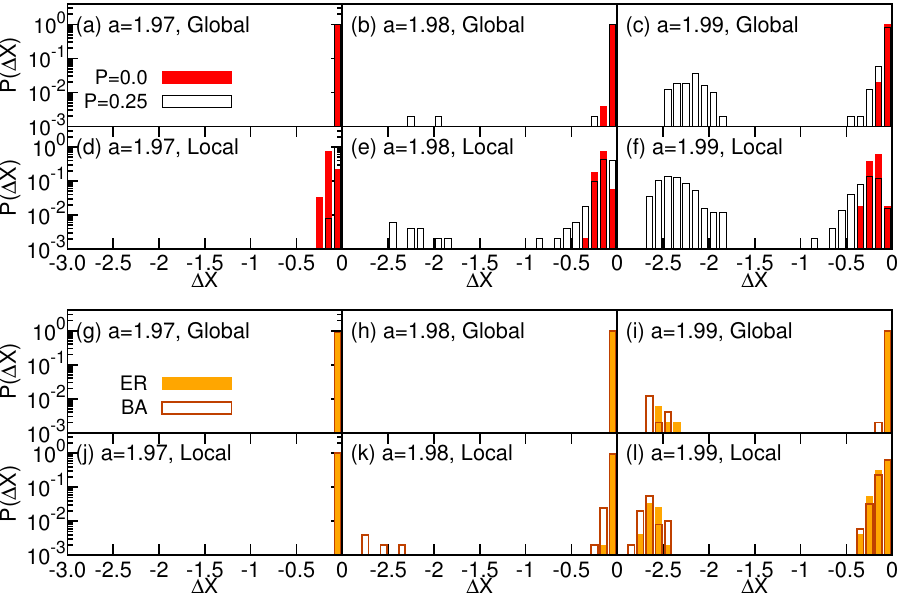}}
\caption{(color online). Distributions of state changes $\Delta X=X_A-X_B$ in the networks of the grazing systems after perturbations. We obtain stationary states $X_{B}$ of a network for given stress $a$ and the initial condition $X_0$ randomly chosen from a uniform distribution $[4.0,6.0]$ and assigned equally to all the nodes. If $X_{B}>0$ (i.e., the network is active), then we give a small shock $S$ to the network and obtain the stationary state $X_{A}$ after the perturbation. (a)-(c) Global perturbations ($S_G=-0.2$) were given to all the nodes in the Watts-Strogatz (WS) networks with $P=0$ and $P=0.25$. (d)-(f) Local perturbations were given to randomly selected $10\%$ of the nodes in the WS networks with $P=0$ and $P=0.25$. (g)-(i) Global perturbations ($S_G=-0.2$) were given to all the nodes in the Erd\H{o}s-R\'{e}yni (ER) network and the Barab\'asi-Albert (BA) network. (j)-(l) Local perturbations were given to randomly selected $10\%$ of the nodes in the ER and the BA networks. For the local perturbations, we randomly selected
10\% of the nodes and set their state to 0.
} \label{fig:SM6}
\end{figure}

\begin{figure}[!tpb]
\centerline{\includegraphics[width=0.99\linewidth,angle=0]{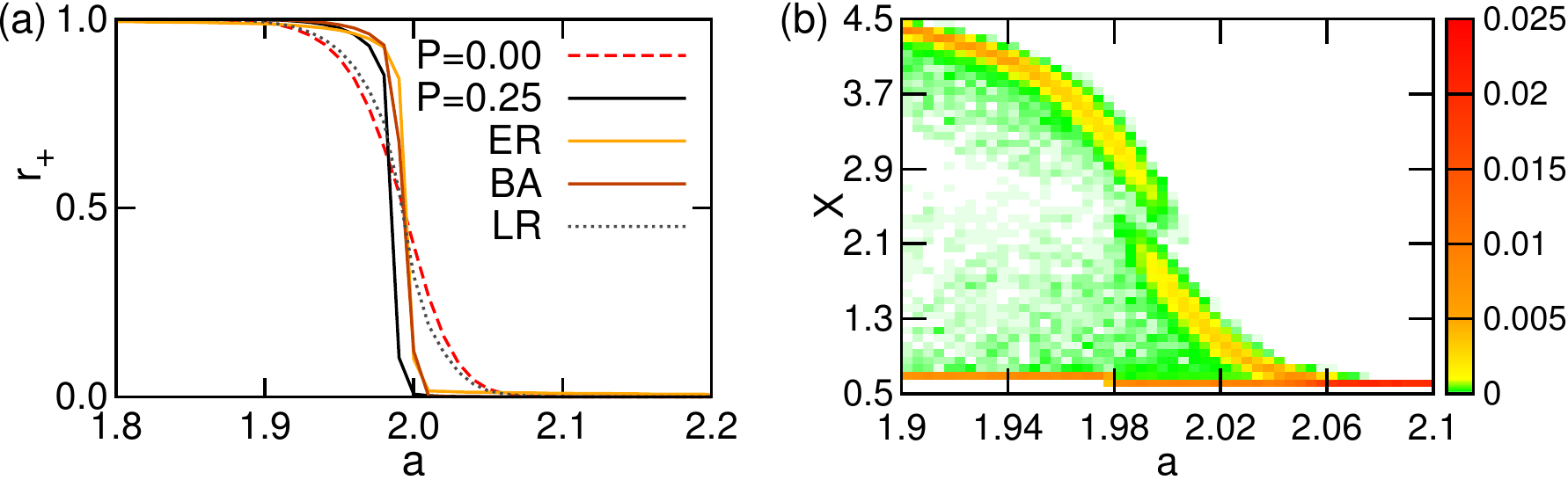}}
\caption{(color online). Suppressed collapse of individual nodes in the networks of the grazing systems. Note that we consider the grazing system $i$ as active when $x_i>1.5$. (a) The ratio of active nodes $r_+$ as a function of the stress $a$ shows the synchronized collapses of the nodes in the Watts-Strogatz (WS) network with $P=0.25$, the  Erd\H{o}s-R\'{e}yni network, and the Barab\'asi-Albert network but the gradual collapses of the nodes in the WS network with $P=0$ and the locally rewired (LR) networks. To generate the LR network, we divide a WS network into four equivalent sublattice before link rewiring and rewire only the links between the nodes in one of the sublattice. Thus the same number of links is rewired in the LR network as in the WS network with $P=0.25$. Each line is averaged over 50 realizations. (b) The stationary state distribution of the LR network shows a small gap indicating the collapse of the rewired region in the upper attractor. We generate $5 \times10^4$ stationary states with the stress $a$ and the initial condition $X_0$ randomly selected from uniform distributions $[1.9,2.1]$ and $[0.1,5.1]$, respectively.} \label{fig:SM_GrSup}
\end{figure}

We consider the grazing system~\cite{NoyMeir1975,May1977}, a paradigmatic model for logistically growing vegetation under grazing stress, for the behavior of the nodes. The state of node $i$ in a network of $N$ grazing systems is evolved by
\begin{equation} \label{eq:S3}
 \frac{dx_i}{dt} = x_i(1-\frac{x_i}{b_i})-\frac{ax_i^2}{x_i^2+1} + \frac{D}{k_i}\sum_{j=1}^{N} A_{ij}(x_j-x_i),
\end{equation}
where $x_i(\geq0)$ is the biomass density of node $i$, $a$ is the maximum consumption rate (i.e., grazing stress), $b_i$ is the carrying capacity of node $i$ randomly drawn from a uniform distribution $B=[5.0,10.0]$, and $D$ is the overall coupling strength ($D=0.3$). The first term represents the logistic growth of the vegetation, the second term represents the grazing stress, and the third term represents the dispersion of the biomass. The network's state $X$ is defined as $X=\frac{1}{N}\sum_i x_i$. We consider the networks with $N=10^4$ and $\langle k\rangle=4$. The bifurcation diagram of the grazing systems when they are isolated is depicted in Fig.~\ref{fig:SBDGr}. We consider the grazing system $i$ as active (collapsed) for $x_i>1.5$ ($x_i<1.5$). In Fig.~\ref{fig:SM4} we consider a deteriorating (ameliorating) scenario where the stationary value of $X(a)$ is obtained by increasing the stress $a$ with an increment (decrement) $\delta a=0.01$ from $a=1.6$ ($a=2.3$) to $a=2.3$ ($a=1.6$). The information on the other numeric experiments is given in the caption of each figure. Figure~\ref{fig:SM5} represents the stationary state distributions of the networks of the grazing systems. Figure~\ref{fig:SM6} depicts the distributions of state changes $\Delta X=X_A-X_B$ in the networks of the grazing systems after perturbations. Figure~\ref{fig:SM_GrSup} shows the suppressed collapse of individual nodes in the networks of the grazing systems.

\begin{acknowledgments}
We thank Hang-Hyun Jo, Seung Ki Baek, and Seung-Woo Son for a careful reading of the preliminary manuscript and for helpful comments. Y.-H.E. acknowledges support from the University of Strathclyde, funding from Ministerio de Econom\'{i}a y Competividad (Spain) through project FIS2016-78904-C3-3-P, and support from the Universidad Carlos III de Madrid, the European Union's Seventh Framework Programme for research, technological development and demonstration under grant agreement (no. 600371), el Ministerio de Econom\'{i}a y Competividad (COFUND2014-51509) and Banco Santander. 
\end{acknowledgments}

\bibliographystyle{apsrev}

\end{document}